\documentclass[12pt]{iopart}
\usepackage{graphicx}
\usepackage{cite}

\usepackage{bm}
\usepackage{mathrsfs}
\usepackage{amssymb}
\usepackage{stmaryrd}

\newcommand{\eq}[1]{(\ref{#1})}
\newcommand{\fig}[1]{figure \ref{#1}}
\newcommand{\be}[1]{\begin{equation}\label{#1}}
\newcommand{\ee}{\end{equation}}
\renewcommand{\vec}{\mathbf}

\usepackage{graphicx}

\usepackage{mathrsfs}

\begin{document}

\title{Total quadruple photoionization cross section of Beryllium in a quasiclassical framework}
\author{Agapi Emmanouilidou$^{1,2}$}
\address{$^{1}$ITS, University of Oregon, Eugene, Oregon 97403-5203\\
$^{2}$Max Planck Institute for the
Physics of Complex Systems, N\"{o}thnitzer Str. 38, 01187 Dresden, Germany } \date{\today}
\begin{abstract}
In a quasiclassical framework, we formulate the quadruple ionization by single photon absorption of the Coulomb five-body problem. We present the quadruple photoionization total cross section of the ground
state of Beryllium for energies up to 620 eV. Our quasiclassical results for energies close to threshold
are in agreement with the Wannier threshold law for four electron escape. In addition, the agreement of our results with a 
shape formula provides support for the overall shape of our total quadruple cross section. Finally, we find that the photon energy where the maximum of the total photoionization cross section occurs for single, double, triple and quadruple photoionization of H, He, Li and Be, respectively, seems to follow a linear relation with the threshold energy for complete
break-up of the respective element.  \end{abstract}
\maketitle   
\section{Introduction}
Multiple ionization of atoms by single photon absorption is a process of great interest since
it probes the correlated motion of many electrons. Quadruple photoionization of the ground state
of Beryllium is the most fundamental atomic process involving four
bound electrons. No other theoretical or experimental study is, to our knowledge, currently available regarding escape of four initially bound electrons. 

 The ionization processes
involving two electron escape can now be treated quite accurately and significant progress 
has been made concerning their understanding \cite{Briggs}. 
 Regarding the three electron escape by single photon absorption from the ground state of Li,
 significant advances have been made over the last few years in obtaining total triple ionization cross sections both experimentally \cite{Wehal98, Wehal00} and theoretically \cite{Pattard,Pindzola04,Emmanouilidou1}. However, still great challenges remain, particularly concerning
 differential cross sections \cite{Malcherek, Colgan,Emmanouilidou2,Emmanouilidou3}. Triple ionization of the ground state of Li by electron impact, that is, a four electron escape process, has been addressed in very recent years. These experimental \cite{Huang} and theoretical \cite{Ghosh} studies address very high energies of the impacting electron.
 
\section{Quasiclassical formulation of quadruple ionization of Beryllium} 
Our construction  of the initial phase space density
$\rho(\gamma)$ for the quadruple photoionization
of Beryllium is similar to the double photoionization of He \cite{RostTobias} and the
triple photoionization formulation of Li which has been detailed in \cite{Emmanouilidou1}.  We
formulate the quadruple photoionization process from the Be ground state
($1s^{2}2s^{2}$) as a two step process in the following way:
\begin{equation}
\sigma^{4+}=\sigma_{abs}P^{4+}
\label{eq:cross}
\end{equation}
where $\sigma_{abs}$ is the total photoabsorption cross section and $P^{4+}$ is the probability 
for quadruple ionization. For $\sigma_{abs}$ we use the experimental data from ref \cite{Atomic}.
Our formulation accounts for the second step.
First, one electron absorbs the photon (photo-electron) at time 
$t=t_{\rm abs}=0$.  Through electronic correlations, the energy is redistributed, 
resulting in four electrons escaping to the continuum.   We first assume that the photon is absorbed by a $1s$-electron at the
nucleus ($\vec r_{1}=0$). This latter approximation becomes exact in the
limit of high photon energy \cite{Kabir}.  For Be, the cross section for photon
absorption from a $1s$ orbital is much larger than from a $2s$ orbital \cite{Rostsem}.
 Hence, we can safely assume that the photo-electron
is a $1s$ electron which significantly reduces the initial phase space
to be sampled.  We also neglect antisymmetrization of the electrons in the initial state. We denote the photo-electron
by 1, the other $1s$ electron by 2 and the two $2s$ electrons by 3 and 4, respectively.
 Following photon absorption, we model the initial phase space
distribution of the remaining three electrons, $1s$ and two $2s$, by the
Wigner transform \cite{Wignertran} of the corresponding initial wavefunction $\psi({\bf
r}_{1}=0,{\bf r}_{2},{\bf r}_{3},{\bf r}_{4},)$, where ${\bf r}_{i}$ are the
electron vectors starting at the nucleus.  We approximate the initial
wavefunction as a simple product of hydrogenic orbitals
$\phi^{\mathrm{Z}_{i}}_{i}(\vec r_{i})$ with effective charges
$Z_{i}$, to facilitate the Wigner transformation.  The $Z_{i}$ are
chosen so that they reproduce the known ionization potentials $I_{i}$, namely
for one of the two 2s electrons $Z_{4}=1.656$ ($I_{4}=0.343\,$a.u.), for the other 2s electron
$Z_{3}=2.314$ ($I_{3}=0.669\,$ a.u.), and for the 1s electron $Z_{2}=3.363$ ($I_{2}=5.656\,$a.u.)(We use atomic units
throughout the paper if not stated otherwise.)  The excess energy,
$E$, is given by $E=E_{\omega}-I$ with $E_{\omega}$ the photon energy
and $I=14.67$ a.u.\ the Be quadruple ionization threshold energy.
Given the above considerations, the initial phase space density
is given by
 \begin{equation}
\label{eq:distribution}
\rho(\gamma) = \mathscr{N}
\delta(\vec{r}_1)\delta(E_{1}+I_{1}-\omega)\prod_{i=2,3,4}W_{\phi^{\mathrm{Z}_{i}}_{i}}
(\vec r_{i},\vec p_{i})\delta(E_{i}+I_{i})
\end{equation}
with normalization constant $\mathscr{N}$.

   We  determine the quadruple ionization probability $P^{4+}$ formally through
\begin{equation}
P^{4+}=\lim_{t\rightarrow\infty}\int_{t_{\mathrm{abs}}}^{t}{\rm d}\Gamma_{\mathcal
{P}^{4+}}\,
\exp((t-t_\mathrm{abs})\mathscr{L}_{\mathrm{cl}})\rho(\Gamma).
\label{eq:intphas}
\end{equation}
The projector $\mathcal P^{4+}$ indicates that
we integrate only over those parts of phase space that lead to quadruple
ionization. $\mathscr{L}_{\mathrm{cl}}$ is the classical Liouville operator
which is defined by the Poisson bracket \{H, \}, with H the
Hamiltonian of the system.  In our case H is the full Coulomb
five-body Hamiltonian.  
 We propagate the electron
trajectories using the classical equations of motion (Classical Trajectory Monte Carlo method \cite{CTMC1,CTMC2}).
 Regularized coordinates \cite{regularized} are used to avoid problems
with electron trajectories starting at the nucleus.  
 We
label as quadruply ionizing those trajectories where the energies of all four electrons are positive, $E_{i}>0$ with $i=1,2,3,4$, asymptotically in time. 

\section{Results}
\begin{figure}
\scalebox{0.5}{\includegraphics{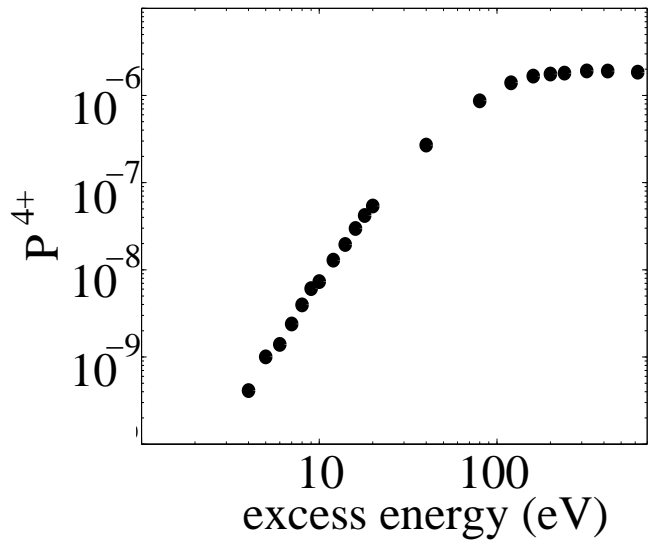}}
\caption{\label{fig:prob}Quadruple photoionization probability as a function of excess energy.}
\end{figure}

\begin{figure}
\scalebox{0.5}{\includegraphics{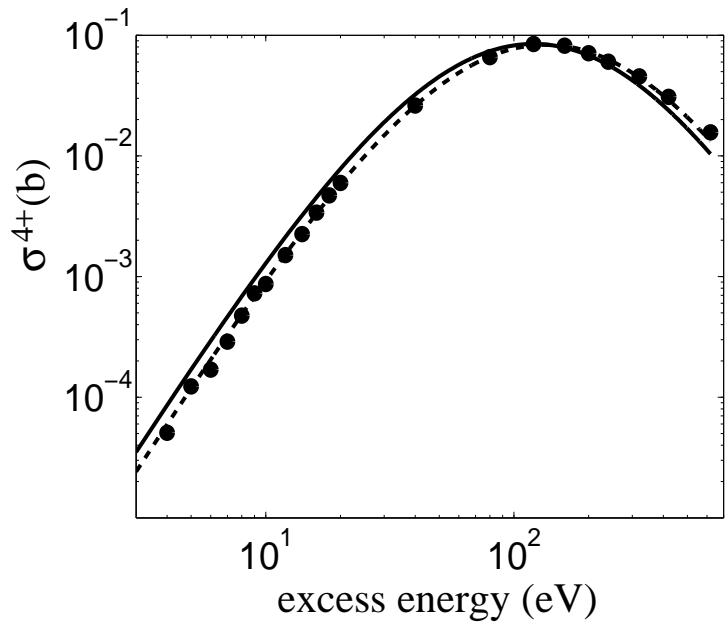}}
\caption{\label{fig:cross}Quadruple photoionization cross section as a function of excess energy. The black circles are our results. For the solid and dashed lines see the main text.}
\end{figure}

 In \fig{fig:prob}, we present our results for  $P^{4+}$. The very small value of the quadruple
 photoionization probability clearly shows the difficulties involved in obtaining theoretical results or experimental measurements of the five-body break-up process. Using \eq{eq:cross} we present in \fig{fig:cross} the results for the quadruple photoabsorption  
cross section (black circles). Since there are no other results currently available we can not
compare our results on an absolute scale. Regarding the accuracy of our results, the standard
relative error is $\propto 1/\sqrt{N_{quad}}$ \cite{standard} with $N_{quad}$ the number of quadruply ionizing events. For energies above 10 eV the $N_{quad}$ we obtain are a few thousands rendering
our results very accurate while for energies from 4eV up to and 9 eV, $N_{quad}$ are at least 1000.

However, a comparison of our results with a shape function applicable to single-photon multiple ionization processes \cite{Shape} provides some indication of how 
well our results reproduce the shape of the total cross section.
This shape function reproduces, by construction, the correct behavior of the single photon multiple-ionization cross section for excess energies close to threshold (Wannier law \cite{Wannier}) and for large excess energies. It depends
on two parameters,  the position $E_{M}$ and height $\sigma_{M}$ of the cross section maximum and
is given by:
\begin{equation}
\label{eq:shape}
\sigma(E)=\sigma_{M}x^{\alpha}(\frac{\alpha+7/2}{\alpha x+7/2})^{\alpha+7/2},
\end{equation} 
where $\alpha$ is the characteristic exponent in the Wannier threshold law and $x$ is the excess energy $E$ scaled by $E_{M}$. For the break up of a four electron 
atom this characteristic exponent was found to be 3.331 for a plane configuration where
the electrons escape on the apexes of a square and 3.288 for a three dimensional configuration where
the electrons escape on the apexes of a tetrahedron \cite{Kuchiev}. From the above two values for $\alpha$ it is
the smallest one, $\alpha=3.288$, that dominates the threshold behavior and thus the value we use
in the shape formula in \eq{eq:shape}. Using the shape formula without any fitting parameters
but substituting for $E_{M}=120$ eV and $\sigma_{M}=0.0847$ b which are the location and height
of the maximum cross section from our results we obtain the solid curve shown in \fig{fig:cross}. Given
that there are no fitting parameters the agreement between our numerical results and the shape formula
is quite good, providing support that the shape of the cross section we obtain is the correct one.
 The agreement
 becomes even better if instead we use the $E_{M}$ and $\sigma_{M}$ as fitting parameters to fit our results with the shape formula. Doing so, we obtain the
 values $E_{M}=134.5$ eV and $\sigma_{M}=0.0834$ b and the agreement of the fitted shape formula
 (dashed curve in \fig{fig:cross}) with our results is indeed very good.  
 
 Also, for small values of the excess energy we have fitted our results for the quadruple
 cross section  with the formula 
 \begin{equation}
 \label{eq:threshold}
 \sigma(E_{\omega})\propto (E_{\omega}/I-1)^{\alpha}
 \end{equation}
 with I the fragmentation threshold energy of beryllium and $E_{\omega}$ the photon energy. For energies from 4 eV to 9 eV our results
 yield a characteristic exponent $\alpha$ of 3.17 which is quite close to the analytical value of 3.288.
 The difference can probably be attributed to the fact that our current results go only down to 4 eV.
  It is a well known fact that as the energy increases the value of $\alpha$ decreases \cite{Emmanouilidou1}.
  Thus, most probably, if we were able to obtain results for energies below 4 eV $\alpha$ would have been closer to the actual value of  3.288. Currently however the numerical difficulties involved in obtaining results below 4 eV are prohibitive.       

   At this point, a few comments are in order. Our quasiclassical formulation is most accurate
 close to threshold. It can describe the process accurately up to at most
 intermediate excess energies while it can not account for high energies. For high
 energies the process can only be described quantum mechanically. In the case of triple ionization
 from the ground state of Li our simple product initial state has proven to be very good in describing
 not only the total cross section \cite{Emmanouilidou1} but also double energy differential cross sections
 up to excess energies of 220 eV, see \cite{Emmanouilidou3}, with $I=203.5$ eV for the triple escape
 process. How good the choice of a product initial state for the case of beryllium is remains to be
 seen when more theoretical and experimental results become available and a comparison with
 our results can be made for a range of excess energies.
           
\begin{figure}
\scalebox{0.4}{\includegraphics{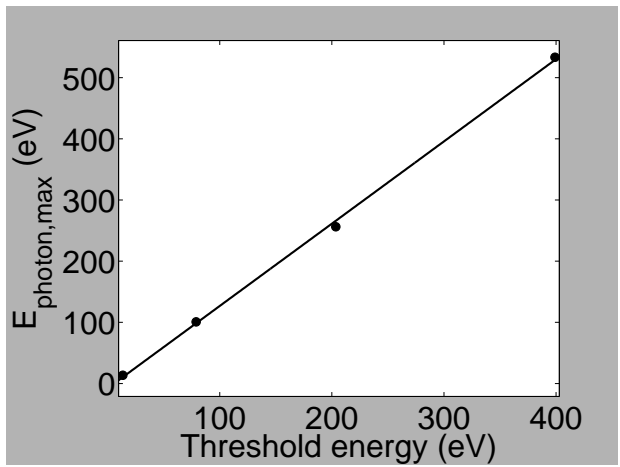}}
\caption{\label{fig:threshold} Photon energy where the maximum of the cross
section for single photoionization of H, double photoionization of He, triple photoionization of Li, and
quadruple photoionization of Be occurs as a function of the threshold energy of the respective process}
\end{figure}

A comparison of the maximum of the single photoionization cross section for H,  $6.3 \times 10^{6}$ b, with the maxima 
for double photoionization of He, $8.76\times 10^3$ b  \cite{Samson}, triple photoionization of Li, $7.47$ b, and quadruple photoionization of Be, 0.0847 b, clearly shows how rare the quadruple
photoionization process is.
For the double ionization cross section of He we use the results presented in ref \cite{Samson}
while for the triple photoionization of Li we use the results presented in ref \cite{Emmanouilidou1}. 
 Another, interesting feature is the relation between the ionization threshold
energy and the photon energy $E_{\omega}=I+E_{M}$ where the maximum of the photoionization cross section of the respective  process occurs. For the single photoionization of H, $E_{M}=0$ eV, for the double photoionization of He, $E_{M}=21.8$ eV \cite{Samson}, while for the triple and quadruple
photoionization of Li and Be $E_{M}$  is respectively 52.7 eV and 134.4 eV. 
The values of 52.7 eV and 134.4 eV were found by fitting \eq{eq:shape} to our results for triple \cite{Emmanouilidou1} and quadruple photoionization, respectively.  
  In \fig{fig:threshold}, we plot $E_{\omega}$ as a function of $I$ for the above four processes. Interestingly, we find
  that for these four processes the ratio of $E_{\omega}$/$I$ is constant. Whether there is a physical reason underlying this constant ratio is an interesting question for future research.  

A future study of interest for the four electron escape in Be would be to investigate whether the collision processes
the four electrons follow to escape to the continum can still
be described as a sequence of three-body subsystems as is the case for Li \cite{Emmanouilidou2}. Another interesting aspect to be investigated is the shape of the inter-electronic angular
distributions for energies close to threshold. In the case, of the three electron escape from the ground state of Li
we found that in an energy regime where the characteristic exponent $\alpha=2.16$ \cite{Klar} is recovered
the angular distribution has a surprising T-shape distribution. This T-shape is different than what one would expect
from the three electrons escaping in the apexes of an equilateral triangle for excess energy
zero \cite{Klar}. At zero excess energy, the four electrons escape in the apexes of a tetrahedron  \cite{Kuchiev}. It remains to be seen whether a surprising pattern will be found for the four electron escape as was the case for the three electrons.

In conclusion, we have presented the first results for the total cross section for the quadruple photoionization of Be. With no other theoretical or experimental results currently available we hope
that the current results will serve as a benchmark calculation for future studies of the Coulomb five-body problem.

{\it{Acknowledgment}} The author is grateful to J.M. Rost, P. Wang, and to T. Pattard for helpful discussions and a  critical reading of the manuscript. She is also indebted to Y. Smaragdakis without whose computational
expertise the current work would not have been possible.


\begin{thebibliography}{19}
\bibitem{Rostsem}
 One can show that the total absorption cross
 section for Be from a 1s orbital is approximately 22 times larger than the one from a 2s orbital
at zero excess energy using a simple semiclassical formula, see Rost J M, J. Phys. B, \textbf{28}, L601 (1995). In addition, 
  assuming the photon absorption happens at the nucleus for a 1s or a 2s orbital the $P^{4+}$
  for quadruple ionization is larger in the case the photon is absorbed from a 1s orbital. The reason is
  that when the photoelectron starting from the nucleus is a 1s it has to ionize one 1s and two 2s electrons
  while if the photoelectron starting from the nucleus is a 2s it has to ionize two 1s and one 2s electron.
\end{thebibliography}
\end{document}